\documentclass[twoside]{article}
\usepackage{Proc_NTSE_18}

\begin{document}
\thispagestyle{plain}
\publref{Karmanov-NTSE-proc}

\begin{center}
{\Large \bf \strut
Bound states of relativistic nature
\strut}\\
\vspace{10mm}
{\large \bf 
V.A.~Karmanov$^{a}$, J.~Carbonell$^{b}$ and H.~ Sazdjian{$^b$}}
\end{center}

\noindent{
\small $^a$\it Lebedev Physical Institute, Moscow, Russia} \\
{\small $^b$\it Institut de Physique Nucl\'eaire, Orsay, France}

\markboth{
V.A.~Karmanov, J.~Carbonell and H.~ Sazdjian}
{
Bound states of relativistic nature} 

\begin{abstract}
Bethe-Salpeter equation, for massless exchange and large fine structure constant $\alpha>\pi/4$, in addition to the Balmer series,  provides another (abnormal) series of energy levels which are not given by the Schr\"odinger equation. So strong field can be created by a
 point-like charge $Z>107$. The nuclei with this charge, though available,
 they are far from to be point-like that weakens the field. Therefore, the abnormal states of this origin hardly exist.

We analyze the more realistic case of  exchange by a massive particle when the large value of coupling constant is typical for the strong interaction. It turns out that
this interaction still generates a series of abnormal relativistic states.
The properties of these solutions are studied. Their existence in nature seems possible.
\\[\baselineskip] 
\end{abstract}

\section{Introduction}

\label{intro}
The Bethe-Salpeter (BS) equation  \cite{bs} is  a relativistic counterpart of the Schr\"odinger equation. In the spinless case, for a two-body system,  it reads
\begin{equation}\label{bs}
\Phi(k,p)=\frac{i^2}{\left[(\frac{p}{2}+k)^2-m^2+i\epsilon\right]\left[(\frac{p}{2}-k)^2-m^2+i\epsilon\right]}
 \int \frac{d^4k'}{(2\pi)^4}iK(k,k',p)\Phi(k',p),
\end{equation}
$p$ is the total four-momentum, $k$ is the relative one. 
The bound state mass sqared is $M^2=p^2=(2m-B)^2$ and $B$ is the (positive) binding energy.
For exchange by the particle with mass $\mu$ the kernel in (\ref{bs}) has the form
\begin{equation}
\label{ladder}
iK(k,k',p)=\frac{i(-ig)^2}{(k-k')^2-\mu^2+i\epsilon}.
\end{equation}
Soon after its derivation, the BS equation was studied by Wick \cite{Wick} and Cutkosky \cite{Cutkosky} in the model of two spinless particles interacting by 
a massless scalar  exchange ($\mu=0$),   since known as  Wick-Cutkosky model. 
Solving equation (\ref{bs})  in the limit of small binding energies ($B/m<<1$), these authors reproduced  the Coulomb spectrum, i.e. the Balmer series 
\begin{equation}\label{Balmer}
B_n=\frac{\alpha^2m}{4n^2} 
\end{equation}
given also by the Schr\"odinger equation with the potential $V(r)=-\frac{\alpha}{r}$, where $\alpha=g^2/(16\pi m^2)$.  (Wick and Cutkosky \cite{Wick, Cutkosky} used other definition of the coupling constant:  $\lambda=\frac{\alpha}{\pi}$).
According to Ref. \cite{Cutkosky}, when $\alpha \to 2\pi$, the ground state mass $M$ determined by eq. (\ref{bs}) tends to 0. When 
$\alpha> 2\pi$, there is no physical solution for the ground state: $M^2$ becomes negative.

To solve the BS equation, Cutkosky has represented the BS
amplitude, for S-wave states, in the following integral form
\begin{equation}\label{Phi}
  \Phi_n(k,p)=\sum_{r=0}^{n-1}\int_{-1}^1\frac{g_{n}^r(z)dz}
      {[m^2-\frac{1}{4}M^2 -k^2-p\cdot k\,z-\imath\epsilon]^{2+n}},
      \quad n=1,2,\ldots\ .
\end{equation}

After substituting (\ref{Phi}) into (\ref{bs}) and performing some
transformations, one finds that the functions $g_{n}^r(z)$ satisfy coupled
integral equations, among which $g_{n}^0$ satisfies a decoupled 
homogeneous equation. The other functions $g_{n}^r$, for $0<r\le (n-1)$,
are then determined from $g_{n}^0$ through the remaining equations.
Denoting henceforth $g_{n}^0$ by $g_n$, one obtains the equation for $g_n$
\begin{equation}\label{eq1}
g''_n(z)+\frac{2(n-1)z}{(1-z^2)}g'_n(z)-\frac{n(n-1)}{(1-z^2)}g_n(z)
+\frac{\alpha}{\pi}\frac{1}{(1-z^2)(1-\eta^2+\eta^2z^2)}g_n(z)=0
\end{equation}
where $\eta=\frac{M}{2m}=1-\frac{B}{2m}$ and the boundary conditions are $g_n(\pm 1)=0$.
For a given $n$, this homogeneous equation has another infinite spectrum
$M_{nk}$, distinct from the ordinary relativistic generalization of the
Balmer series, corresponding to bound states $g_{nk}$ with binding
energies $B_{nk}=2m-M_{nk}$, depending on a second integer quantum
number $k=1,2,3,\ldots$. In the limit of small binding energies,
$B_{nk}$ is independent of $n$,  namely:
\begin{equation}\label{Bk}
B_{nk} \approx B_k=m\exp\left(-\frac{2\pi^{3/2}k}{\sqrt{\alpha-\pi/4}}\right), 
\end{equation}
For $k=0$ and arbitrary $n$, the levels are still given by the Balmer
series (\ref{Balmer}), corresponding to the so-called normal ones.  
The abnormal solutions $g_{nk}(z)$  have $k$ nodes in $z$. 
For the even values of $k$ the solutions $g_{nk}(z)$ are symmetric  in $z\to -z$ and for odd $k$'s  are antisymmetric.  
Corresponding BS amplitude in the rest frame is symmetric or antisymmetric relative to $k_0\to -k_0$. In \cite{cia} is was shown that the antisymmetric solutions do not contribute  to the $S$-matrix and therefore they are hardly observable.  Therefore we will  consider the symmetric (normal and abnormal) states only.                                                                                                                                                      

To summarize,  for massless exchange, in addition to the Balmer series,  the BS equation predicts for each $n$, another series of  states with binding energies $B_{nk}$ given  -- in the limit 
$B/m<<1$ -- by eq. (\ref{Bk}). 
These states exist only if $\alpha >\frac{\pi}{4}$. Their binding energies tend to zero when $\alpha\to \frac{\pi}{4}$. 
They are absent in the spectrum of non-relativistic Schr\"odinger equation and therefore they were called "abnormal". 
 
 
\begin{figure}[!ht]
 \centerline{\includegraphics[width=0.8\textwidth]{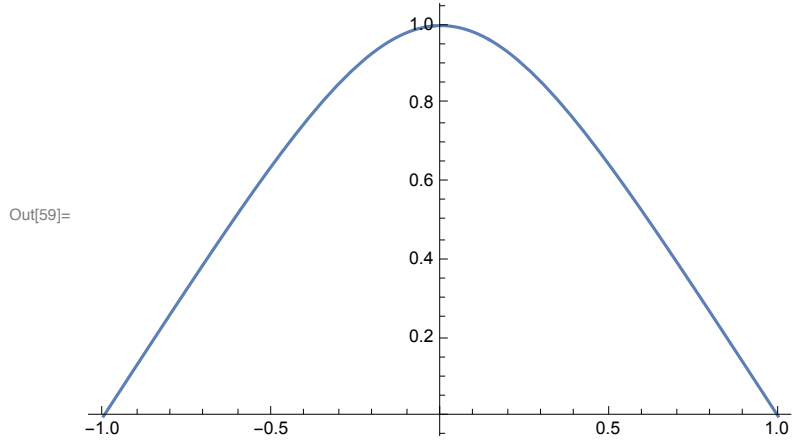}}
\caption{Normal solution $g(z)$ of eq. (\ref{eq1}) ($\mu=0$), corresponding to $n=1$, $k=0$, $B=0.2$, $\alpha=1.786$.}
\label{fig1}       
\end{figure}

\begin{figure}[!ht]
 \centerline{\includegraphics[width=0.8\textwidth]{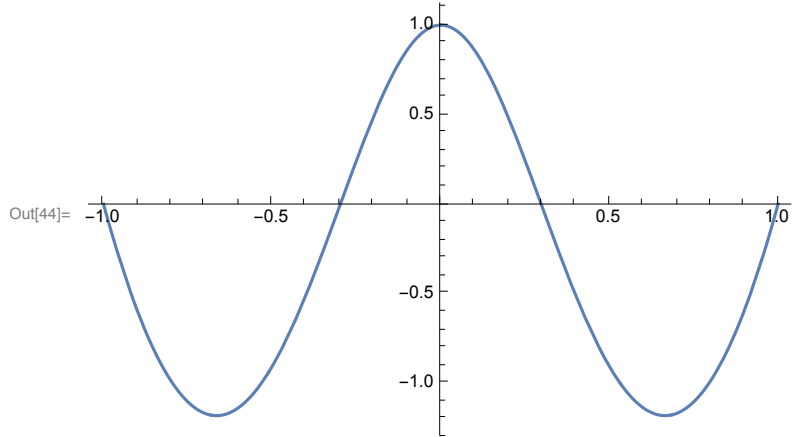}}
\caption{Abnormal symmetric solution  $g(z)$  of eq. (\ref{eq1}) ($\mu=0$), corresponding to $n=1$, $k=2$, $B=0.2$, $\alpha=17.19$.}
\label{fig3}       
\end{figure}

In the limit $\eta=\frac{M}{2m}\to 1$ Wick and Cutkosky  found analytical solutions of eq. (\ref{eq1}). For arbitrary $\eta$ we solved this equation  numerically.
The examples of normal and abnormal symmetric solutions $g(z)$ are shown in Figs. \ref{fig1} and \ref{fig3} respectively.  These both solutions correspond to $n=1$ and differ from each other by the $k$-values: $k=0$ and $k=2$. 

The aim of our research is to answer the question:  can the abnormal states exist in the nature or not? In the case of the massless exchange considered by Wick and Cutkosky and sketched above, the answer seems to be negative. The required coupling constant 
$\alpha>\frac{\pi}{4}$ is too large to be reached in practice. 
Indeed, since the value $\alpha=\frac{1}{137}$ corresponds to  $Z=1$, $\alpha>\frac{\pi}{4}$  corresponds to the charge $Z> \frac{\pi}{4}/(\frac{1}{137})\approx 107$. 
Nuclei with this and larger charge, though  do not exist in nature, were created in a laboratory  ($Z=107$ corresponds to bohrium). However, they are far from to be point-like.  Since charge is distributed in a large volume, the strength of the electric field is reduced. Therefore, to create an abnormal state,  one needs even larger (maybe, much larger) value of $Z$. This makes the problem unrealistic. 

However, the value $\alpha=\frac{\pi}{4}\approx 0.78$  is normal when dealing with strong interactions. 
The latters are modeled by exchange by a massive particle exchange. 
Therefore, in our research we will replace the massless exchanged particle by the massive one and we will study whether  or not the abnormal states still survive. 
It turns out that  the most peculiar properties of the abnormal states in the WC model (existence of the critical coupling constant $\alpha_c=\frac{\pi}{4}$, determining the existence of the abnormal states; simultaneous appearance and disappearance of infinite series of levels,  when the coupling constant crosses the critical value) are a consequence of the zero exchanged mass. 
One could then expect that these properties do not exist anymore in the massive case. 
However, this does not forbid the existence of the abnormal states at all, though a definite answer requires some research.

\section{Non-zero exchanged mass}\label{solv}

For solving this problem, it is still convenient  to use an integral representation for the BS amplitude similar to (\ref{Phi}). 
Namely:
\begin{equation}\label{Phi1}
\Phi(k,p)=\int_0^{\infty}d\gamma\int_{-1}^1\frac{g(\gamma,z)dz}{\left[\gamma+m^2-\frac{1}{4}M^2 -k^2-p\cdot k\,z-\imath\epsilon\right]^3}
\end{equation}
This representation has been proposed by Nakanishi \cite{nakanishi}.
For simplicity of notations, we omit the indices $n,k$.
In contrast to (\ref{Phi}) for the massless case, the weight function
$g(\gamma,z)$ in (\ref{Phi1}) depends on an additional variable
$\gamma$ and, correspondingly, the integral (\ref{Phi1}) is double. 
The massless exchange corresponds to the particular situation where
the function $g(\gamma,z)$ can be expressed, concerning its
$\gamma$-dependence, as a combination of a delta function and $(n-1)$ of
its derivatives in $\gamma$:
\begin{equation}\label{delta}
g(\gamma,z)=g_n(\gamma,z)=\sum_{r=0}^{n-1} \delta^{(r)}(\gamma)
g_n^r(z),\quad n=1,2,\ldots\ .
\end{equation}
Inserting (\ref{Phi1}) in the BS equation (\ref{bs}), one can derive
an equation for $g(\gamma,z)$.  
Some properties of solutions will be still studied analytically whereas the
spectrum and corresponding solutions will be found numerically.

For the ladder BS kernel, though, in a little bit complicated form, the equation for the  weight function $g(\gamma,z)$ was firstly derived in  \cite{KusPRD95}.
For arbitrary BS kernel, though in the form containing integrals in both sides of equation, the equation for $g(\gamma,z)$ was derived in 
\cite{bs1}. It reads:
\begin{equation}\label{bs1}
\int_0^{\infty}\frac{g(\gamma',z)d\gamma'}{\Bigl[\gamma'+\gamma +z^2 m^2+(1-z^2)\kappa^2\Bigr]^2}
=\int_0^{\infty}d\gamma'\int_{-1}^{1}dz'\;W(\gamma,z;\gamma',z') g(\gamma',z'),
\end{equation}
where $\kappa^2-m^2-\frac{1}{4}M^2$

In the canonical form 
\begin{equation}\label{fsv}
g(\gamma,z)=\int_0^{\infty}d\gamma'\int_{-1}^{1}dz'\;{\mathcal V}(\gamma,z;\gamma',z') g(\gamma',z'),
\end{equation}
for the ladder BS kernel, the equation for $g(\gamma,z)$ was derived in \cite{FrePRD14}. In this work, the expression for the kernel 
$V(\gamma,z;\gamma',z')$, corresponding to the ladder BS kernel, was found. 

At last, in ref. \cite{CKF_PLB_2017} it was noticed that the l.h.-side of eq. (\ref{bs1}) is the generalized Stieltjes transform which can be inverted analytically. In this way, the equation for $g(\gamma,z)$ in the canonical form, valid for arbitrary BS kernel, was derived. For the ladder BS kernels, the kernels $V(\gamma,z;\gamma',z')$ in eq. (\ref{fsv}), found in \cite{CKF_PLB_2017} and \cite{FrePRD14}, coincide with each other.  Useful research of the non-relativistic limit of the BS equation was done in ref. \cite{wand}.

\vspace{1cm}

\begin{figure}[!h]
 \centerline{\includegraphics[width=0.8\textwidth]{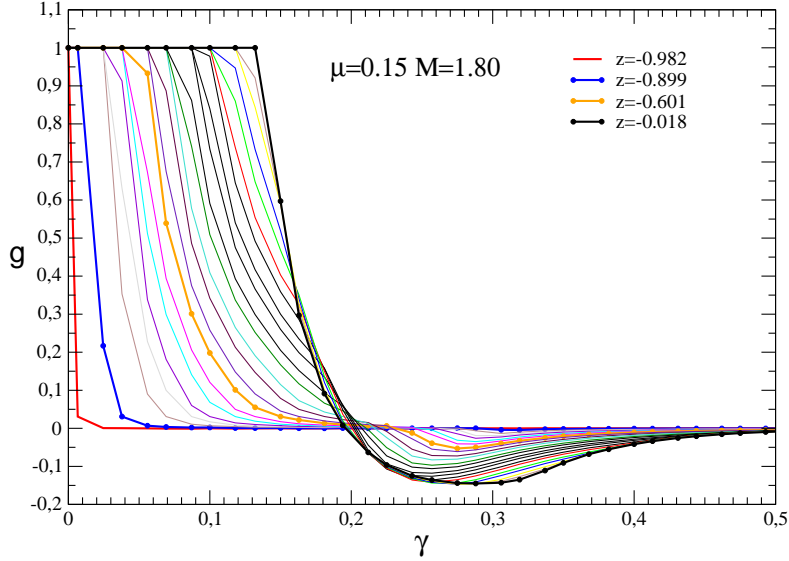}}
 \bigskip
\caption{Nakanishi weight function $g(\gamma,z)$ corresponding  $m=1$, $\mu=0.15$, $B=0.2$ and $\alpha=2.1$ as a function of $\gamma$ for a fixed $z$.}
\label{Fig_g}       
\end{figure}

We will analyze the equation in the form (\ref{fsv}) with the kernel, derived in  \cite{FrePRD14} from the ladder kernel (\ref{ladder}). This kernel reads: 
 \begin{equation}\label{Vf}
{\mathcal V}(\gamma,z;\gamma',z') = + \frac{\alpha m^2}{2\pi}\times
\left\{
\begin{array}{ll}
\displaystyle{h(\gamma,-z;\gamma',-z')}, &\quad \mbox{if $-1\le z'\le z\le 1$} \\
\displaystyle{h(\gamma,z;\gamma',z')}, &\quad  \mbox{if $-1\le z\le z'\le 1$}
\end{array}
\right.
\end{equation}
with the following function $h$
\begin{equation}\label{hp}
h(\gamma,z;\gamma',z')=  \theta(\eta) \; P( \gamma,z,\gamma',z' )   +  Q(\gamma',z' ),
\end{equation}
where
\[ P(\gamma,z,\gamma',z' )  =   \frac{B}{\gamma A\Delta}\frac{ 1+z }{( 1+z' )}  -  C(\gamma,z,\gamma',z' )   \]
with
\begin{eqnarray*}
&&A (\gamma',z' )                         =   \frac{1}{4}{z'}^2 M^2+\kappa^2+\gamma', \quad
B(\gamma,z,\gamma',z' )         = \mu^2+\gamma'-\gamma\frac{1+z'}{1+z},   \cr
&& C(\gamma,z,\gamma',z' )        =  \int_{y_-}^{y_+}\chi(y)dy,
\quad
Q(\gamma',z' )                         =  \int_{0}^{\infty}\chi(y)dy ,\cr
&&\Delta (\gamma,z,\gamma',z' ) = \sqrt{B^2-4\mu^2A}.
\end{eqnarray*}
The functions $C$ and $Q$ contain the function
$$
\chi(y) = \frac{y^2}{\left[y^2+A+y(\mu^2+\gamma')+\mu^2\right]^2}   \label{chi}
$$
and the integration limits in $C$ are given by
$y_{\pm} = \frac{-B\pm\Delta}{2A}.$
The argument $\eta$ of the $\theta$-function in the first term of (\ref{hp}) is:
$$
\eta=-B-2\mu\sqrt{A}=\gamma\frac{1+z'}{1+z}-\mu^2-\gamma'-2\mu\sqrt{\frac{1}{4}{z'}^2 M^2+\kappa^2+\gamma'}.
$$

The results of solving  numerically equation (\ref{fsv}) with the parameters $\mu=0.15, B=0.2$ are displayed in Figure \ref{Fig_g}. 
The coupling constant is $\alpha=2.1$ and corresponds to a "normal" state.
They have been obtained in a recent work \cite{SKC_2019} by using the same spline techniques than in \cite{bs1}.
The  Nakanishi weight function $g$ has been computed by several authors in the past either by solving
eq. (\ref{bs1}) or equivalently in its normal form (\ref{fsv}).
None of them put in evidence the striking behavior
of this quantity  manifested in Fig. \ref{Fig_g}, that is
a step-like function on variable $\gamma$ and flat behaviour in some domain in variable $z$.
The difficulty is due, on one hand, to the numerical instabilities on $g$ related to the $\epsilon$-trick introduced in 
Ref.  \cite{bs1} and, on other hand, to the difficulty to obtain a flat behavior using a Gaussian-like basis expansion as done in Refs. \cite{FrePRD14,FSV_EPJC75_2015,VFS_FBS56_2015,PdP_FBS57_2016}. 
This  behavior has been also proved analytically in \cite{SKC_2019}  and will be discussed below in Sec. \ref{prop}.
\vspace{1cm}

\begin{figure}[!h]
 \centerline{\includegraphics[width=0.8\textwidth]{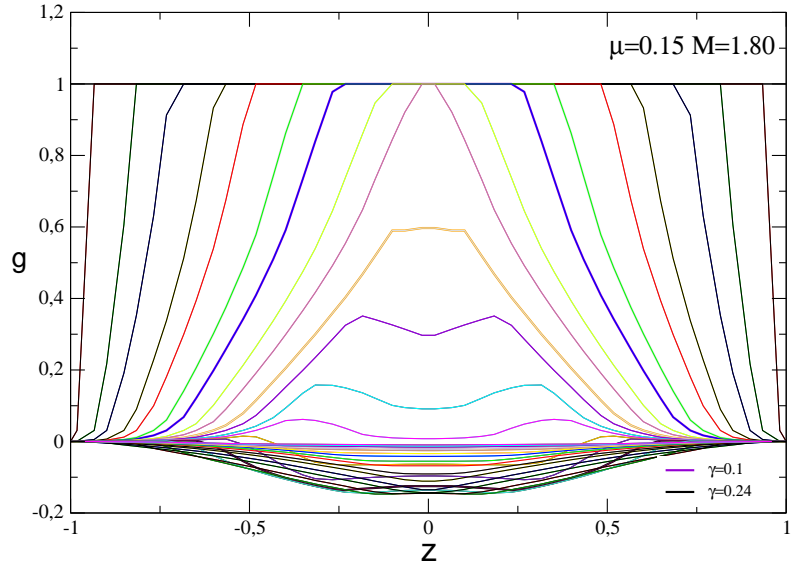}}
\caption{The same as in Fig. \ref{Fig_g}  as a function of $z$ for a fixed $\gamma$.}
\label{Fig_g_a}       
\end{figure}


\section{Non-relativistic limit}\label{non-rel}

The "relativistic world" differs from the non-relativistic one by existence of the limiting value of speed of any object or signal, which is 
identified with the speed of light $c$. Calculating via a relativistic equation the binding energy, corresponding to a normal state, and  taking the limit $c\to\infty$, we should obtain the non-relativistic binding energy.  
The abnormal states, not existing in the non-relativistic limit, should disappear when $c\to\infty$. The relativistic equations presented above implied $c=1$. To study the limit $c\to\infty$, we should now restore the speed of light $c$ in these equations. 

The strategy is the following. We should introduce $c$ in the parameters which consist the input of equation. This, of course, automatically makes influence on the parameters which are "output" (found from the equation), therefore we should leave their value untouched.  This means that we should replace $m$ by  $mc^2$. As for the total mass, it is not an independent parameter (not an input), but it is expressed as 
$M=2mc^2-B$. Therefore one {\it should not} put $M\to Mc^2$. The same is valid for $B$, since it also is not input, but it is found from the equation (already containing $mc^2$) as an eigenvalue. Therefore one also {\it should not} put $B\to Bc^2$, but one should keep the binding energy$B$ as it is. Since in QED the coupling constant is $\alpha=e^2/(\hbar c)$, the $c$ value should appear explicitly. Therefore $\alpha$ should be replaced by $\alpha/c$. 

In this replacement, there was no any reference to the mass of exchange particle. Therefore it is valid not only in QED, but also in the Yukawa model  with exchange by a massive particles.  A subtle point is the replacement of the exchanged mass $\mu$. The ladder exchange results in the Yukawa potential with the factor $\sim \exp(-\mu r)$. Restoring $c$ in this factor, we get: 
$\exp\left(-\frac{\mu c^2 r}{\hbar c}\right)$. In the limit $c\to\infty$ we find the zero-range potential. The Yukawa potential shrinks into a delta-function. 
However, in the present research, we study, how the energies found from a relativistic equation itself turn into the energies determined by the Schr\'odiger equation with given potential $V(r)$ and we are not interested in the effects resulting from the variation of $V(r)$ itself v.s. $c$. The shrink is avoided if we put $\mu\to\mu c$, not $\mu\to\mu c^2$.

\begin{figure}[!h]
 \centerline{\includegraphics[width=0.8\textwidth]{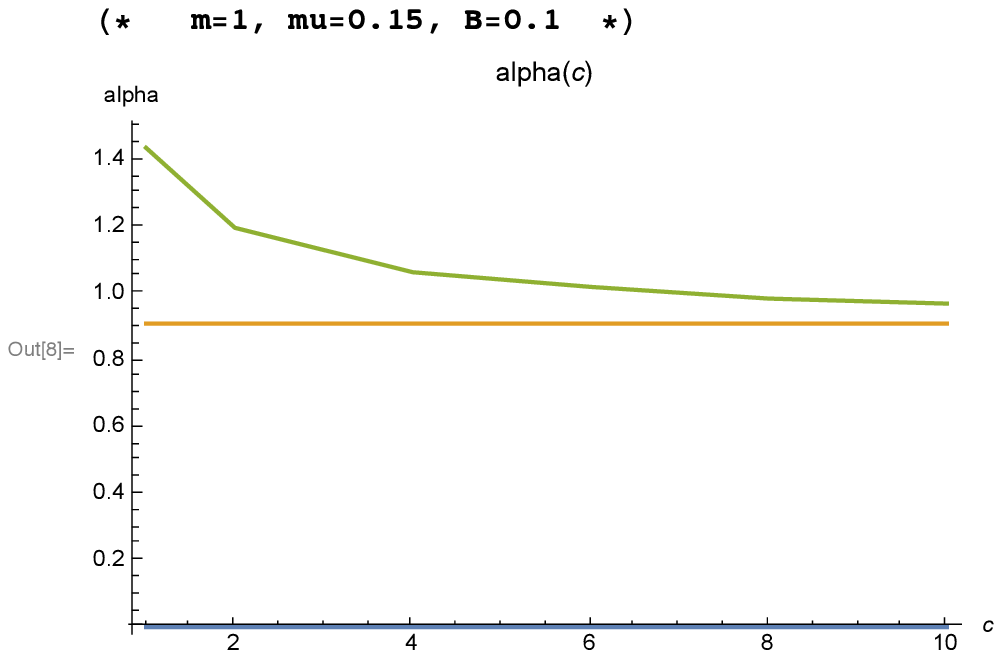}}
\caption{The coupling constant $\alpha$ vs. speed of light $c$  for the parameters $m=1$, $\mu=0.15$, $B=0.1$.
\newline
Decreasing line is found from the relativistic equation (\ref{fsv})  for the "normal" solution; horizontal  line is the limiting value of the decreasing line, the coupling constant in the Yukawa potential in the Schr\"odinger equation. }
\label{fig_alpha}       
\end{figure}

\begin{figure}[!h]
 \centerline{\includegraphics[width=0.8\textwidth]{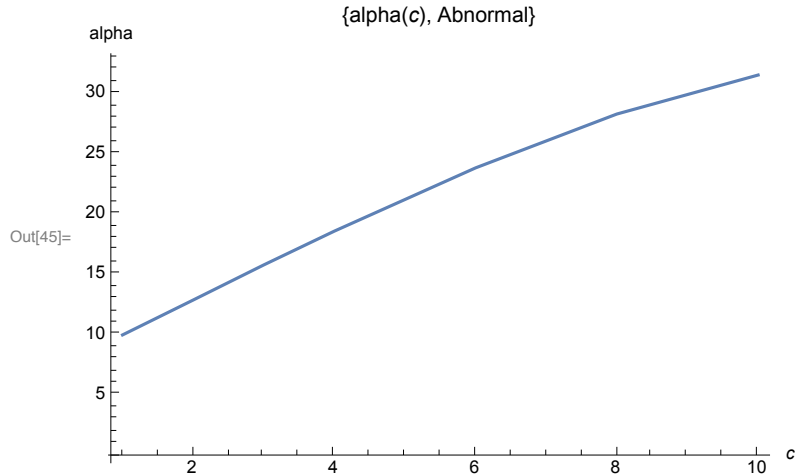}}
\caption{The coupling constant $\alpha$ vs. speed of light $c$  for the parameters $m=1$, $\mu=0.15$, $B=0.1$  for the "abnormal" solution.}
\label{fig_alpha_a}       
\end{figure}

Making these replacements in the kernel ${\mathcal V}$ of the equation (\ref{fsv}), we solved the latter numerically and, varying $c$, we studied the behavior of two energy levels. More precisely, for fixed binding energy $B$ we studied the behavior of the coupling constant $\alpha$ vs. $c$ in the interval $1\leq c\leq 10$. We put $m=1$, $\mu=0.15$ and $B=0.1$.  For one of the states, which we associate with the "normal" solution shown in Fig. \ref{fig_alpha}, we found $\alpha(c=1)\approx 1.45$. For another state, which we associate with the "abnormal" solution Fig. \ref{fig_alpha_a}, we found $\alpha(c=1)\approx 10.$ The results for $\alpha(c)$ for these two states are shown in 
Figs. \ref{fig_alpha} and \ref{fig_alpha_a}. These curves  have opposite behavior vs. $c$ (decrease and increase).

\bigskip

\bigskip

At Fig. \ref{fig_alpha} we see that in the non-relativistic limit ($c\to\infty$) $\alpha$ decreases and tends to a limiting finite value $\alpha \approx 0.9$. This value is just the coupling constant in the Yukawa potential providing the binding energy $B=0.1$ in the Schr\"odinger equation. Therefore we associate this solution with the "normal" one, having the non-relativistic limit. The decrease $\alpha$ vs. $c$ seen at Fig. \ref{fig_alpha} can be easy explained qualitatively. As it was noticed in many works, the relativistic effects, been added to the non-relativistic dynamics, result in an effective repulsion. Therefore, when we go in the non-relativistic limit  ($c$ increases), we decrease this repulsion. Hence, we need smaller coupling constant $\alpha$ to keep the fixed value of the binding energy $B=0.1$.   

According to the curve $\alpha(c)$ shown in Fig. \ref{fig_alpha_a}, the value $\alpha(c)$ increases when $c$ increases, at least in the interval $1\leq c\leq 10$. 
 The disappearance of the abnormal states  when $c$ increases means that the corresponding energy levels are "pushed out" into the continuum spectrum. That is, they move up and cross the value $B=0$. To prevent this movement and to keep these levels on the place, say, at $B=0.1$, like in it is at Fig. \ref{fig_alpha_a} for $c=1$, one should increase the attraction. Hence, when $c$ increases, we need larger coupling constant $\alpha$, what is observed at Fig. \ref{fig_alpha_a}. Therefore we associate this solution with the "abnormal" one.

These results demonstrate the existence of the abnormal states in the solution of the BS equation with the massive ladder kernel (we assume that the qualitative behavior of $\alpha(c)$ vs. $c$ can be extrapolated for larger $c$). At least, one of them is found and corresponding $\alpha(c)$ is shown at Fig. \ref{fig_alpha_a}.

Coming back to the zero-mass exchange, we can put in the binding energy $B_k$, eq. (\ref{Bk}), $m\to mc^2$ and $\alpha\to\alpha/c$. Then, solving this equation relative to $\alpha$ we find:
\begin{equation}\label{alB}
 \alpha=c\left(\frac{\pi}{4}+\frac{4\pi^3k^2}{\log^2\frac{B_k}{mc^2}}\right)
 \end{equation}
This formula gives an analytical example of the dependence $\alpha(c)$ which has no finite limit at \mbox{$c\to\infty$}, that serves as an undoubted property of abnormal state.

We remind that, for the massless exchange, there exists another criterion  to select an abnormal state: it is existence of nodes of the solution $g(z)$ (see  Fig. \ref{fig3}). Though  only one of these two criterions is enough to distinguish the abnormal states, it is useful to establish both.
For the massive exchange, we discussed so far only one criterion  to select an abnormal solution: absence of a finite limit of $\alpha(c\to\infty)$. Below we will formulate another criterion, also based on analysis of nodes of the solution $g(\gamma,z)$. 

\section{Properties of the $z$-dependence of the solution $g(\gamma,z)$}\label{prop}
\begin{figure}[!h]
 \centerline{\includegraphics[width=0.8\textwidth]{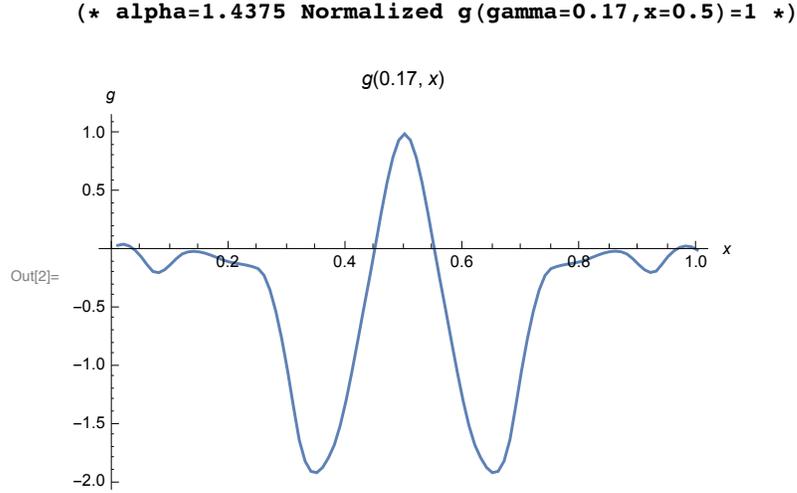}}
\caption{Normal solution $g(\gamma=0.17,x)$ of eq. (\ref{fsv})    with the kernel (\ref{Vf}) vs.  $x=\frac{1}{2}(1+z)$ for the parameters $m=1$, $\mu=0.15$, $B=0.1$, $\alpha=1.4375$ for fixed value $\gamma=0.17$.}
\label{fig5}       
\end{figure}

As discussed in Sec. \ref{intro} for massless exchange, the normal solution $g(z)$ has no nodes (see Fig. \ref{fig1}). 
However, for massive exchange, this property cannot be used to distinguish a normal solution.  
In Fig. \ref{fig5} we show the solution $g(\gamma,x)$ (with the parameters $m=1$, $\mu=0.15$, $B=0.1$, $\alpha=1.4375$) 
for fixed value $\gamma=0.17$ vs. $x=\frac{1}{2}(1+z)$. Instead of $z$, we introduced for convenience new variable $x$ varying in the limits $0\leq x\leq 1$. This is a normal solution since the dependence $\alpha(c)$, corresponding to this solution, is just  shown in 
Fig. \ref{fig_alpha}.
\bigskip

\begin{figure}[h]
 \centerline{\includegraphics[width=0.8\textwidth]{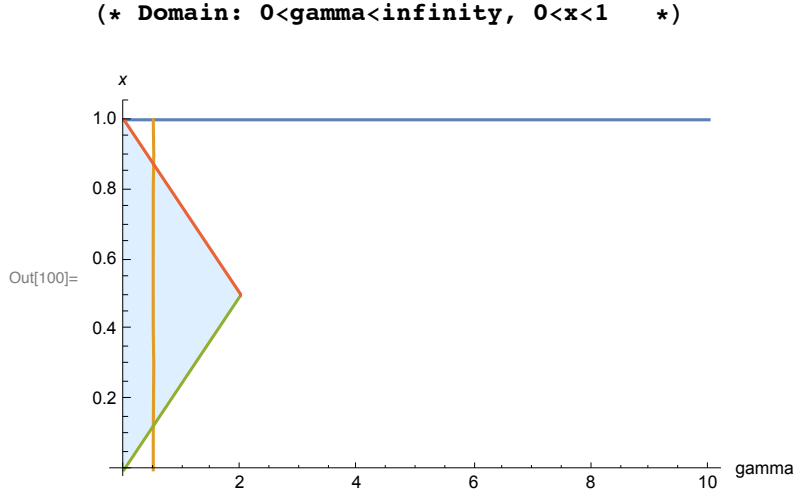}}
\caption{Domain of definition of the function $g(\gamma,x)$. Shadowed area is the domain defined by eq. (\ref{dom}) where  $g(\gamma,x)=const$.}
\label{fig6}       
\end{figure}

In spite of the fact that it is normal, it has nodes vs. $x$ for a fixed $\gamma$. However, it turns out, that the behavior of $g(\gamma,x)$
in different parts of the domain of its definition is qualitatively different. There exists an area where  $g(\gamma,x)=const$.  This area is defined (up to a factor) by:
\begin{equation}\label{dom}
 \gamma \leq \gamma_0(x)\sim \left\{
 \begin{array}{cc}
 \mu x,& \mbox{if $0\leq x\leq\frac{1}{2}$}
 \\
 \mu (1-x),& \mbox{if $\frac{1}{2}\leq x \leq 1$}
 \end{array}
 \right.
\end{equation}
The domain ($0\leq \gamma <\infty$, $0\leq x\leq 1$) is shown in Fig. \ref{fig6}, whereas the shadowed  area corresponds to eq. (\ref{dom}).
\bigskip

\begin{figure}[!h]
 \centerline{\includegraphics[width=0.8\textwidth]{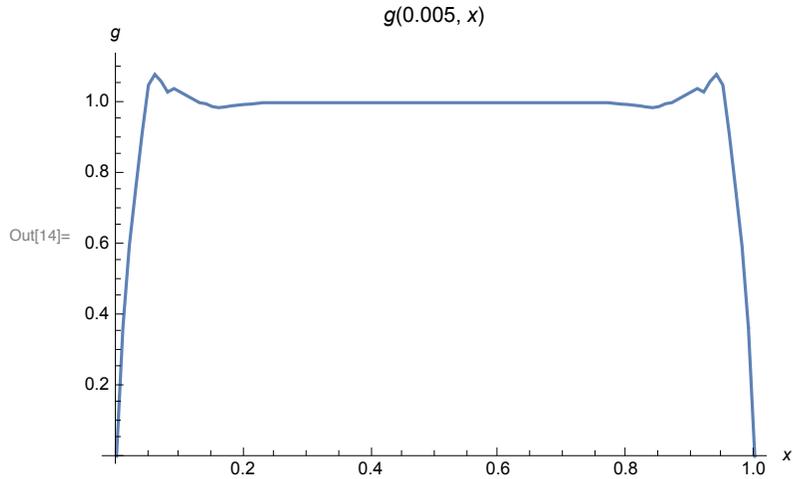}}
\caption{Normal solution $g(\gamma=0.005,x)$ of eq. (\ref{fsv})    with the kernel (\ref{Vf}) vs.  $x$ for the same parameters as in Fig. \ref{fig5}.}
\label{fig7}       
\end{figure}

\begin{figure}[!h]
 \centerline{\includegraphics[width=0.8\textwidth]{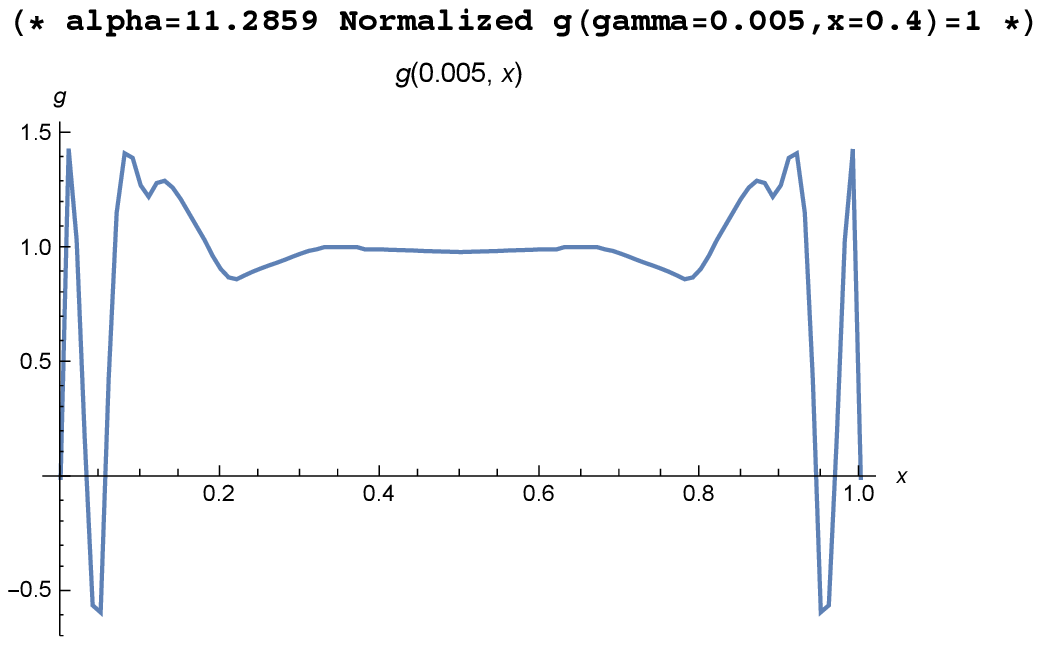}}
\caption{Abnormal symmetric solution $g(\gamma=0.005,x)$, $\alpha=11.2859$ of eq. (\ref{fsv})  with the kernel (\ref{Vf}) vs.  $x$ for the same parameters as in Figs. \ref{fig5} and \ref{fig7}}
\label{fig9}       
\end{figure}

Let us fix the $\gamma$-value, taking it from the area eq. (\ref{dom}) (we take $\gamma=0.005$) and consider the $x$-dependence of $g(\gamma=fixed,x)$.
This corresponds to variation of $x$ along the vertical line crossing the triangle in Fig. \ref{fig6}. Solving numerically eq. (\ref{fsv}), for the normal solution, which was shown for $\gamma=0.17$, we now obtain Fig. \ref{fig7}.
We see that in the interval $x_1<x<x_2$, where the vertical line in Fig. \ref{fig6} is inside the shadowed triangle,  the function $g(\gamma=0.005,x)$ vs. $x$ is indeed constant, as expected. Outside this interval, when $x$ is close to 0 or to 1, $g(\gamma=0.005,x)$ varies. However, it has no nodes. In this respect, the behavior of the normal solution $g(\gamma=0.005,x)$ is analogous to the behavior of  normal $g(z)$ for massless exchange. The latter has nodes nowhere (except for the points $x=0,1$, where the nodes are imposed by the boundary conditions). 

The symmetric abnormal solution for the same parameters (though, of course, for different binding energies) are shown in Fig. \ref{fig9}. It is still constant when $x$ is inside of the domain defined in (\ref{dom}) and they have the nodes outside this domain, in contrast to the normal solutions. Like for the massless exchange, this gives us another criterion (except for the limit 
$\alpha(c\to\infty)$) to distinguish, in the case of massive exchange, the abnormal solutions from the normal ones.

We emphasize that these results are based on the numerical calculations. It would be useful to derive them analytically.

\section{Conclusion}\label{concl}

Like the Dirac equation predicts antiparticles,  the BS equation predicts  bound states having a purely relativistic origin:
these are the so called "abnormal" states, not given by the Schr\"odinger equation.  

We have found that these states, previously found in the Wick-Cutkosky model (scalar massless exchange), 
requiring for their existence a large coupling constant, exist also for the interaction provided by massive exchange with values of the coupling constant typical for the strong interaction. 
It is worth conjecturing that  these states  could manifest in some natural processes. One should analyze from this point of view the systems which meet difficulties in describing them as ordinary bound states and requiring exotic speculations. Maybe, some of these systems are "abnormal" ones.

For deeper understanding the abnormal states, it would be useful to calculate corresponding electromagnetic form factors, to compare them with the "normal" form factors and calculate also the transition form factors normal $\leftrightarrow$ abnormal states. 

To clarify the content of "abnormal" state vector (i.e., the contributions into its full norm of the Fock components with different numbers of particles) still remains an intriguing task. The preliminary results look as follows. For the normal state, when binding energy tends to zero, the contribution of the two-body sector dominates in the norm of full state vector. On the contrary, for the abnormal state, when binding energy tends to zero ($\alpha\to\frac{\pi}{4}$), the contribution of the two-body sector  in the norm of full state vector decreases.


\end{document}